\pdfoutput=1
\documentclass[11pt]{article}

\usepackage[final]{acl}

\usepackage{times}
\usepackage{latexsym}

\usepackage{subfig}
\usepackage{multirow}
\usepackage{enumitem}
\usepackage{amsmath,amssymb}
\usepackage{utfsym}
\usepackage{threeparttable}
\usepackage{bm}
\usepackage[switch]{lineno}
\usepackage{booktabs}
\usepackage{graphicx}
\usepackage{CJKutf8}
\usepackage{booktabs}
\usepackage{algorithm}
\usepackage{float}
\usepackage{algpseudocode} 
\definecolor{Gray}{gray}{0.9}
\usepackage[T1]{fontenc}

\usepackage[utf8]{inputenc}

\usepackage{microtype}

\usepackage{inconsolata}

\usepackage{graphicx}
\usepackage{multirow}
\usepackage{enumitem}
\usepackage[utf8]{inputenc}
\usepackage[most]{tcolorbox} 
\usepackage{amsmath}   
\usepackage[most]{tcolorbox} 
\tcbuselibrary{breakable}
\newtcolorbox{promptbox}[1]{
    colback=blue!4!white,           
    colframe=blue!50!white,            
    colbacktitle=blue!50!white,        
    coltitle=white,            
    fonttitle=\bfseries\large, 
    title={#1},                
    arc=10pt,                  
    outer arc=10pt,
    left=10pt,                 
    right=10pt,                
    top=10pt,                  
    bottom=10pt,               
    boxrule=1.5pt,             
    enhanced,
    breakable,
    pad at break=0mm
}
\title{EAGER: Enrich-and-Align Generative Query Recommendation from Clicked Items in E-commerce Search}

\begin{document}
\author{
  Shuwei Yuan\footnotemark[1] \quad
  Mingqian Ding\footnotemark[1] \quad
  Luxin Liu \quad
  Rong Xiao \quad
  Xiaoyi Zeng \\
  Alibaba International Digital Commerce Group \\
 \texttt{\{yuanshuwei.ysw, dingmingqian.dmq, xique.llx\}@alibaba-inc.com} \\
  \texttt{\{xiaorong.xr, yuanhan\}@taobao.com}
}

\maketitle
\renewcommand{\thefootnote}{\fnsymbol{footnote}}
\footnotetext[1]{Equal contribution.}
\renewcommand{\thefootnote}{\arabic{footnote}}
\begin{abstract}

E-commerce platforms increasingly display clickable query suggestions alongside items in the user feed, enabling users to refine or expand their intent without manually reformulating queries. Existing approaches either mine suggestions from historical logs—limited to past behavior and blind to long-tail, personalized intents or rely on off-the-shelf LLMs whose lack of platform-specific knowledge yields fluent but generic queries disconnected from real click behavior. We propose \textbf{EAGER} (\textbf{E}nrich-\textbf{a}nd-\textbf{A}li\textbf{G}n g\textbf{E}nerative Query \textbf{R}ecommendation), a two-stage framework for generating query suggestions from clicked items. In the enrichment stage, supervised fine-tuning (SFT) follows a four-stage curriculum that scales information richness (from item-only to user-conditioned) and reasoning depth (from direct to chain-of-thought). Each stage incorporates rationale augmentation, diversity regularization, and self-distillation. In the alignment stage, we post-train via GRPO with a hybrid reward of multiple rule-based business signals and a preference-aware click reward. Extensive offline experiments and online A/B test demonstrate the effectiveness of EAGER, which has been deployed in production at a major e-commerce platform.

\end{abstract}

\section{Introduction}
\label{sec:introduction}

Query Recommendation~\cite{xu2026aigq,min2025prompting} is a core module in modern e-commerce search that helps users quickly express their shopping intents~\cite{yin2026clicks,chen2025llm}. Beyond traditional query completion~\cite{shao2025great} and suggestion~\cite{guo2026onesug} in the search box, industrial platforms increasingly provide context-aware query recommendation at key user touchpoints~\cite{min2025ctr} to naturally support subsequent exploration in recommendation scenarios.

\begin{figure}[t]
  \centering
 \includegraphics[width=1.0\linewidth]{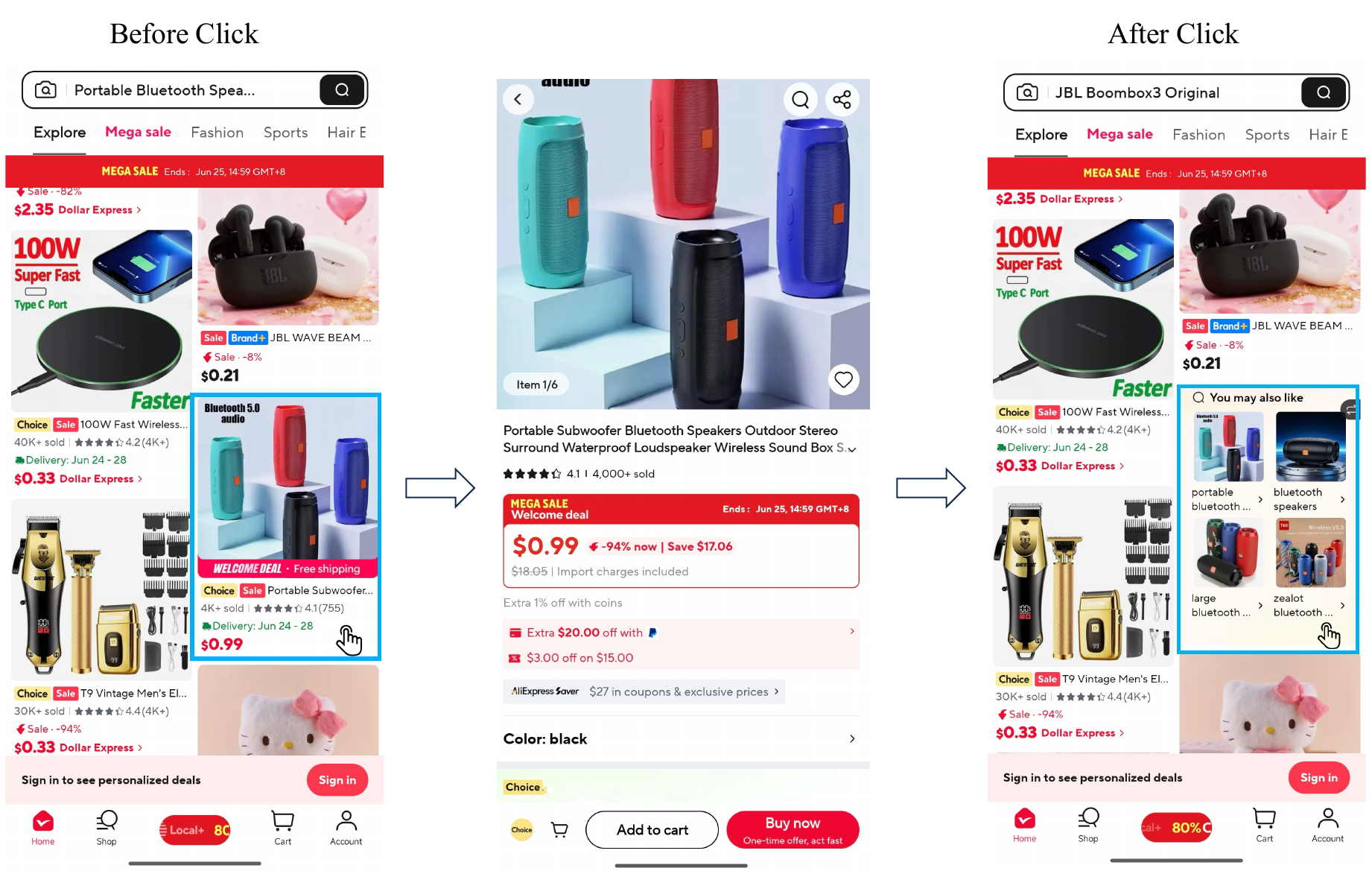}
  \caption{Query recommendation from clicked items. When a user clicks an item in the recommendation feed, the system presents a set of query suggestions to capture their immediate search intent.}
  \label{fig:first}
\end{figure}

Among these touchpoints, a particularly high-value scenario arises when a user clicks an item and returns to the recommendation feed (Figure~\ref{fig:first}): the system presents a set of query suggestions to capture the user's immediate search intent. We refer to this task as item-to-query (I2Q) recommendation throughout this paper.

Existing online methods~\cite{bie2026pushgen} mainly rely on collaborative-filtering-based statistical features~\cite{papadakis2022collaborative}, mining item-query correlations through historical behavior co-occurrence and depending on pre-existing query pools~\cite{si2022model}. They perform well in data-sufficient scenarios but cannot model fine-grained semantics or user-personalized preferences. For long-tail and new items, the recommended queries tend to be biased toward generic head queries and fail to fully capture users' potential intents. Our online traffic logs further reveal a clear positive correlation between query-pool richness and downstream conversion, motivating an approach that expands the candidate space beyond what log-based mining can reach.

Recent advances in large language models~\cite{team2026kimi,zeng2026glm,deepseekai2025deepseekv32pushingfrontieropen} have spurred industrial exploration into generative query recommendation. Existing studies~\cite{sun2026quality,guo2026onesug,bi-etal-2026-relist} have applied LLMs to search intent understanding and query generation, achieving promising results in general query recommendation tasks. However, generating query recommendations from clicked items presents two core challenges. First, a single item click reveals only one explicit choice, yet the system must produce a diverse query set covering the user's multiple latent intents. Vanilla LLMs~\cite{coreteam2026mimov2flashtechnicalreport,yang2025qwen3technicalreport} produce fluent but generic queries without personalization or priority ordering, and cannot bridge this gap. Second, the generated query set must simultaneously satisfy multiple conflicting objectives---real user click-through rates and downstream conversion on the one hand, and deployment constraints such as format compliance, diversity on the other. These heterogeneous objectives span both user preference alignment and business rule enforcement, and cannot be jointly optimized through standard likelihood training alone.

To address these challenges, we propose a two-stage framework. The first stage enriches the query generation space through supervised fine-tuning with multi-source intent supervision, curriculum learning, and self-distillation, enabling the model to cover diverse user intents. The second stage aligns generation with online deployment constraints and user click preferences via GRPO~\cite{Guo_2025} with hybrid rewards combining rule-based and preference-aware click rewards. The two stages are complementary: SFT maximizes intent coverage offline, while GRPO post-training ensures the generated queries meet online deployment criteria without sacrificing the diversity established in the first stage. Our main contributions are summarized as follows:

\begin{itemize}
\item We propose EAGER, a generative I2Q recommendation framework that bridges item recommendation and search in e-commerce by generating personalized query sets from clicked items and user context, deployed in production at a large-scale e-commerce platform.
\item We design a two-stage training framework: SFT enhances intent coverage and query diversity through multi-source supervision, curriculum learning, and self-distillation; GRPO further aligns generation with user preferences and online business constraints via hybrid reward optimization. A shared Preference-Aware Reward Model (PARM) anchors both stages on the same production click signal, ensuring training consistency.
\item Extensive offline evaluations and online A/B tests confirm that our framework substantially improves core production metrics including user engagement and payment conversion.
\end{itemize}

\graphicspath{{./image/}}
\begin{figure*}[t]
  \centering
  \includegraphics[width=\textwidth]{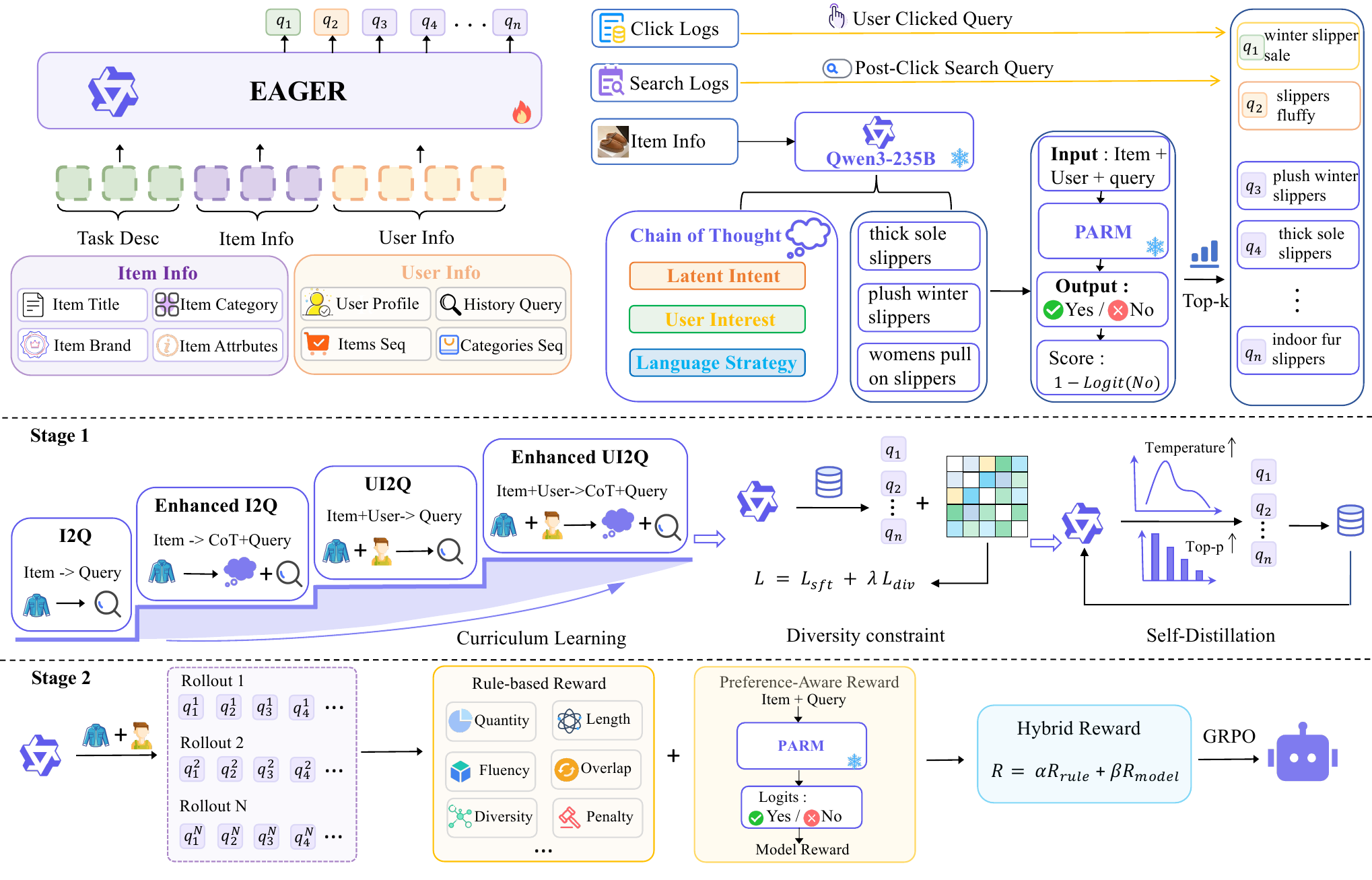}
  \caption{Overview of EAGER. The framework consists of two stages: (1) supervised fine-tuning with multi-source supervision, curriculum learning, and self-distillation for query enrichment; (2) GRPO-based alignment with hybrid rewards for online deployment.}
  \label{fig:overview}
\end{figure*}


\section{Preliminary}
\label{sec:preliminary}

We study user-contextualized generative query recommendation in the I2Q setting. Given a recently clicked item $i$, user context $u$, and online serving constraints $c$, the model aims to generate a query set $Y=\{q_1,\ldots,q_K\}$ of size $K$. Unlike pool-based retrieval, which selects candidates from a predefined query inventory, our task requires the model to generate queries directly from item semantics and immediate user behavior, subject to constraints on accuracy, diversity, and online deployability.

\section{Methodology}
\label{sec:method}

We describe the two stages below: supervised fine-tuning for query enrichment (\S\ref{sec:sft}) and reinforcement learning with hybrid rewards for online alignment (\S\ref{sec:rl}).

\subsection{Supervised Fine-Tuning for Query Enrichment}
\label{sec:sft}
The goal of this stage is to generate a diverse set of queries from the clicked item and user context to cover users' latent search intents and provide a richer candidate space for subsequent reinforcement learning and online ranking. To this end, we design the SFT pipeline from three aspects: supervision construction, task design, and training augmentation.

\subsubsection{Multi-source Intent-ordered Supervision}
\label{sec:multi}
A key challenge in the I2Q task is that a user's real click usually reflects only one explicit choice, and thus cannot fully capture the user's latent intents. Moreover, relying solely on clicked queries leads to sparse supervision. To better mine potential user intents, we construct multi-source intent-ordered supervision labels consisting of three types of queries:
\begin{equation}
    Y = [q^{click}; Q^{post}; Q^{llm}],
\end{equation}
where $q^{click}$ denotes the query the user clicked on the recommendation panel, serving as the strongest supervision signal. $Q^{post}$ denotes the queries actively issued by the user shortly after clicking a recommended query, which further reflects the user's immediate post-click intent. 

To alleviate the sparsity of click logs, we further perform offline query expansion for popular items using a large language model. Specifically, we prompt the LLM with the item title, category, attributes, and possible usage scenarios to generate a set of candidate queries, denoted as $Q^{llm}$.

However, LLM-generated queries do not inherently have reliable priorities. Directly appending them to the labels may cause the model to treat queries of different quality as equally important supervision, thereby introducing noise. To address this issue, we score the LLM-generated queries with the \textbf{Preference-Aware Reward Model (PARM)}—a generative click predictor fine-tuned from Qwen-4B on production click logs—which also serves as the model-based reward in GRPO (\S\ref{sec:clk_model}), unifying supervision selection and reward optimization under the same production preference signal. Training details are in Appendix~\ref{sec:appendix_clk}.

\subsubsection{Rationale-Augmented Curriculum Learning}
After constructing the SFT data, we design a rationale-enhanced curriculum learning strategy. Specifically, we organize training into four tasks with increasing difficulty.

The first task is item-to-query (\textbf{I2Q}), which takes only item information as input and generates item-related queries. This task aims to establish a basic mapping between item attributes and e-commerce search queries. Different from standard I2Q, \textbf{Enhanced I2Q} incorporates rationales generated by a 235B teacher LLM. The model first analyzes the key item attributes, usage scenarios, and possible search directions, and then generates the target \texttt{queries}.

The third task is user-item-to-query (\textbf{UI2Q}), which further introduces personalized user information. This task guides the model to move from generic item-level query generation to personalized query generation. The fourth task is \textbf{Enhanced UI2Q}, which jointly incorporates item information, user context, and rationale supervision. By reasoning over the user's historical preferences and current click behavior, the model learns to generate queries that better match the user's immediate intent.

The overall SFT stage is optimized with the standard autoregressive negative log-likelihood objective:
\begin{equation}
\mathcal{L}_{\text{SFT}} = - \sum_{t=1}^{T} \log p_\theta\bigl(y_t \mid y_{<t}, x\bigr).
\end{equation}

\subsubsection{Diversity-aware SFT Enhancement}
To further improve query diversity, we introduce two SFT enhancement strategies: diversity-regularized SFT and self-distillation.

First, to mitigate the homogenization problem in multi-query generation, we continue fine-tuning the model on a set of high-quality samples and add a diversity regularization term to the autoregressive loss. The objective is defined as:
\begin{equation}
\mathcal{L} = \mathcal{L}_{\text{SFT}} + \lambda \mathcal{L}_{\text{div}},
\end{equation}
where $\lambda$ controls the weight of the diversity loss. The diversity loss penalizes excessive similarity among queries within the same output sequence:
\begin{equation}
\mathcal{L}_{\text{div}} = \frac{1}{K(K-1)} \sum_{i\neq j} \text{sim}(q_i, q_j),
\end{equation}

Second, we introduce a self-distillation mechanism to expand the model's generation space. Specifically, we use the diversity-enhanced model to perform sampled inference on a subset of prompts. During inference, we increase the temperature and $\text{top-}p$ values to encourage the model to generate more diverse candidate queries. We then filter the generated results by removing queries with invalid formats, abnormal lengths, or business rule violations. The remaining high-quality outputs are used to reconstruct additional training data for the next round of SFT.

\subsection{Reinforcement Learning with Hybrid Rewards}
\label{sec:rl}
Although the SFT stage enhances the richness of generated queries, it is fundamentally suited to offline supervision data. To further align the generative model with the online recommendation objectives of the I2Q task, we introduce a GRPO-based reinforcement learning approach with hybrid rewards.

\subsubsection{Verifiable Rule-Based Reward}

The rule-based reward is designed to enforce deterministic business constraints in the I2Q scenario. Given an input $x$ and a set of generated queries
$
Y = \{q_1, q_2, \dots, q_K\},
$
we define this reward as a weighted sum of multiple sub-rewards:
\begin{equation}
R_{\text{rule}}(x, Y) = \sum_{m=1}^{M} w_m r_m(x, Y),
\end{equation}
where $r_m(x, Y)$ denotes the $m$-th rule reward and $w_m$ is its corresponding weight. We design eight verifiable rule rewards mainly from the perspectives of user experience, compliance, and fluency. Detailed definitions are provided in the appendix.

These rewards enforce basic display requirements and business constraints, reducing format errors and redundant generation.

\begin{table*}[t]
    \centering
    \resizebox{\textwidth}{!}{
    \begin{tabular*}{\linewidth}{@{\extracolsep{\fill}}lll*{6}{c}}
        \toprule
         & \textbf{Method} 
        & \multicolumn{2}{c}{\textbf{Hard HR}} 
        & \multicolumn{2}{c}{\textbf{Soft HR}} 
        & \multicolumn{2}{c}{\textbf{Diversity}} 
        & \multirow{2}{*}{\textbf{Rule}}  \\
        \cmidrule(lr){3-4} \cmidrule(lr){5-6} \cmidrule(lr){7-8}
        & 
        & @1 & @all 
        & @1 & @all 
        & APD & Distinct-2 
        & \\
        \midrule
        & Gemini 3.1& 0.42  &  0.52 &  70.04 &  93.08 &  69.46  & 76.59 & 7.83\\
        & GPT-5.2 & 0.32 & 0.58  &  \underline{75.73} & \textbf{94.52} &  67.74 & 77.17 & 8.57\\
        \midrule
        & Qwen3-1.7B & 0.00 & 0.04  &  0.09 & 34.59 & 41.82  & 63.94 & 3.55\\
        & Qwen3-4B & 0.08  &  0.34 &  31.56 &  56.06 & 52.33   & 70.75 & 5.39\\
        & Qwen3-8B &  0.14 &  0.37 &  52.83 & 73.97  & 54.49  & 72.83 &5.72\\
        & Qwen3.5-397B-A17B & 0.08  & 0.10 &  69.08 &87.98 & 64.91 & 76.33  & 8.19\\
        \midrule
        & SFT & 0.33  & 0.67  & 74.04 & 86.85  & 53.05 & 76.79 &7.65\\
        & GRPO &  0.27 & 0.61  &  66.58 &  79.36 & 62.84 & 76.48  & 7.73\\
        & SFT-GRPO & 0.36  & 0.68  & 74.19  & 86.91 & 53.34  & 78.65  &7.88\\
        & EAGER-base & \textbf{0.65}  &  \textbf{1.27} & 74.83  &  93.27 & \underline{72.13} & \textbf{83.75}  &\underline{9.04}\\
        & EAGER & \underline{0.44} & \underline{0.87}  &  \textbf{77.96} & \underline{93.82} & \textbf{72.96}  & \underline{83.04}  &\textbf{9.16}\\
        \bottomrule
    \end{tabular*}
    }
    \caption{Offline Experimental Results. EAGER and its variants are implemented based on Qwen3-1.7B. Best results are in bold and the second-best results are underlined.}
    \label{tab:off}
\end{table*}

\begin{table}[t]
\centering
\begin{tabular}{l l c c}
\toprule
\textbf{Model} & \textbf{HardHR} & \textbf{SoftHR} & \textbf{APD} \\
\midrule
EAGER-1.7B & 0.44  &  77.96 & 72.96 \\
EAGER-4B   & 0.54 &  78.39 & 79.13 \\
EAGER-8B   & 0.56 & 78.65 &  78.84\\
\bottomrule
\end{tabular}
\caption{Experimental results with different Large Language Model backbones.}
\label{tab:size}
\end{table}

\subsubsection{Model-based Reward via PARM}
\label{sec:clk_model}
While rule-based rewards can enforce deterministic constraints, they cannot directly estimate whether a generated query is attractive to a specific user under the current item context. To align the reward signal with real user feedback in the I2Q scenario, we adopt PARM (introduced in \S\ref{sec:multi}) as the model-based reward, which scores each generated query by its predicted click likelihood given the item and user context.

Given an item $i$, user context $u$, and candidate query $q$, PARM estimates the likelihood that the user would click $q$ under the current exposure context. We train the model to generate a natural language label:
$
a \in \{\text{yes}, \text{no}\}$.
The training details are provided in the Appendix.

During inference, we use the probability of the ``no'' label to construct the click reward. Intuitively, if the model is less likely to predict ``no'', the query is more likely to match the user's current click preference. The reward for a query is defined as:
\begin{equation}
r_{\text{model}}(i, u, q) = 1 - p_{\text{PARM}}(\text{no} \mid i, u, q).
\end{equation}

The set-level model reward is computed as the average click reward over all queries:
\begin{equation}
R_{\text{model}}(i, u, Y) = \frac{1}{K} \sum_{k=1}^{K} r_{\text{model}}(i, u, q_k).
\end{equation}

As described in \S\ref{sec:multi}, the same scoring function $1 - p_{\text{PARM}}(\text{no} \mid i, u, q)$ is used to rank LLM-generated candidates during SFT data construction, converting unordered queries into weakly ordered supervision.

\subsubsection{GRPO Policy Optimization}

We combine the two rewards to construct the overall reward function for reinforcement learning:
\begin{equation}
R(x, Y) = \alpha R_{\text{rule}}(x, Y) + \beta R_{\text{model}}(i, u, Y).
\end{equation}
Driven by the hybrid reward, GRPO optimization further guides the model to generate \texttt{query} lists that better satisfy business constraints, align with real user click preferences.

\section{Experiment}
\label{sec:experiment}

\subsection{Experimental Setup}
\paragraph{Metrics.}
Offline, we evaluate along three axes: relevance, diversity, and compliance.
\textbf{Soft HR@$K$} is our \emph{primary} relevance metric: a generated query
counts as a hit if it exactly matches the logged clicked query, or if at least
60\% of its terms appear in the title of the item the user clicked after
entering search. Soft HR is thus grounded in the second-hop item click, which is
the objective of this scenario. \textbf{Hard HR@$K$} requires an exact string
match with the single logged query and is reported only as a diagnostic of
exact-expression reproduction: because I2Q is open-vocabulary and one-to-many, a
single logged query is only one of many valid verbalizations, so a low Hard HR
does not by itself indicate low query quality. Diversity is measured by
\textbf{APD} and \textbf{Distinct-2}, and business compliance by the
\textbf{Rule} score. All HR values are reported in percent. Online, we report
UCTR and PCTR for engagement, and DAC, Pay Count, and L2P for downstream
conversion.
\paragraph{Baselines} 
To evaluate the effectiveness of our method, we compare it with open-source and closed-source LLMs of varying scales and capabilities, using them to assess the generative query recommendation ability of general-purpose LLMs without task-specific adaptation. We also include several model variants under different training and inference settings to provide a comprehensive comparison of recall performance.

\begin{table*}[t]
    \centering
    \resizebox{\textwidth}{!}{
    \begin{tabular*}{\linewidth}{@{\extracolsep{\fill}}lll*{7}{c}}
        \toprule
         & \textbf{Method} 
        & \multicolumn{2}{c}{\textbf{Hard HR}} 
        & \multicolumn{2}{c}{\textbf{Soft HR}} 
        & \multicolumn{2}{c}{\textbf{Diversity}} 
        & \textbf{Rule}\\
        & 
        & @1 & @all 
        & @1 & @all 
        & APD & Distinct-2 
        & \\
        \midrule
        & EAGER &0.44  &  0.87 & 77.96  &  93.82 & 72.96 & 83.04  &9.16\\
        & \textit{w/o} GRPO &  0.36 & 0.64  & 77.45  & 93.76  & 72.84 & 81.77  &7.21\\
        & \textit{w/o} Self-Distill  & 0.41 & 0.65  & 75.60 & 93.42  & 69.62  & 81.73 & 8.53\\
        & \textit{w/o} DIV &  0.38 & 0.56  & 78.34  &  93.49 &  68.24 & 78.86 & 8.42\\
        & \textit{w/o} DIV + Self-Distill & 0.37 & 0.51  &  78.07 & 92.03 & 65.05  & 78.23  & 8.17\\
        \bottomrule
    \end{tabular*}
    }
    \caption{The experimental results of the ablation study.}
    \label{tab:abla}
\end{table*}

\begin{table}[t]
\centering
\begin{tabular}{lcc}
\toprule
\textbf{Category} & \textbf{Metric} & $\boldsymbol{\Delta}$\\
\midrule
\multirow{2}{*}{First}  & UCTR      & +0.83\% \\
                        & PCTR      & +1.28\% \\
\midrule
\multirow{3}{*}{Second} & DAC       & +2.98\% \\
                        & Pay Count & +3.49\% \\
                        & L2P       & +2.38\% \\
\bottomrule
\end{tabular}
\caption{Online Experimental Results of EAGER.}
\label{tab:on}
\end{table}

\subsection{Offline Experiments}
Table~\ref{tab:off} reports the offline results. Closed-source LLMs achieve competitive Soft HR but relatively weak Hard HR, indicating that they can generate semantically related queries yet fail to precisely match real user expressions. Zero-shot open-source models perform poorly across all metrics, confirming the necessity of task-specific training for the I2Q task. We compare several framework variants: \textbf{SFT} uses multi-source supervision without curriculum staging or diversity enhancement; \textbf{GRPO} applies hybrid-reward policy optimization directly on the backbone; \textbf{SFT-GRPO} chains the two; and \textbf{EAGER-base} adds curriculum learning, diversity regularization, and self-distillation atop SFT-GRPO but omits PARM-based query ranking. SFT and SFT-GRPO significantly outperform the backbone, while GRPO alone improves diversity but degrades relevance without SFT initialization. EAGER-base achieves the highest Hard HR and Distinct-2, confirming that curriculum learning and self-distillation expand the generation space. The full EAGER trades moderate Hard HR decrease for gains in Soft HR@1, APD, and Rule score—a favorable trade-off for online deployment where downstream ranking reorders candidates.

Table~\ref{tab:size} shows that EAGER scales well across backbone sizes, with EAGER-4B achieving the highest APD and EAGER-8B the best Soft HR.

\subsection{Ablation Study}
Table~\ref{tab:abla} presents the ablation results. Two observations deserve emphasis. First, removing DIV slightly \emph{raises}
Soft HR@1 (78.34 vs.\ 77.96) while sharply reducing APD (68.24 vs.\ 72.96) and
Distinct-2 (78.86 vs.\ 83.04). This is the expected coverage--diversity
trade-off: concentrating probability mass on the single most likely
verbalization improves exact relevance against the logged query but yields a
homogeneous query set. Since the scenario exposes $K$ queries simultaneously and
a downstream ranker reorders them, conceding 0.38 Soft HR@1 for +4.72 APD and
+4.18 Distinct-2 is favorable, and the online results confirm that the more
diverse set converts better. Second, removing GRPO leaves relevance essentially
unchanged (Soft HR@All 93.76 vs.\ 93.82) but degrades Rule from 9.16 to 7.21 and
Distinct-2 from 83.04 to 81.77, indicating that GRPO's primary role is
business-rule compliance rather than relevance. Removing self-distillation
reduces Hard HR@All from 0.87\% to 0.65\% and APD from 72.96 to 69.62,
confirming that model-generated high-quality data broadens the training
distribution.

\begin{table}[t]
\centering
\setlength{\tabcolsep}{4pt}
\begin{tabular*}{\columnwidth}{@{\extracolsep{\fill}}lcccc@{}}
\toprule
\textbf{Configuration} & \multicolumn{2}{c}{\textbf{Soft HR}} & \textbf{APD} & \textbf{Dist-2}\\
\cmidrule(lr){2-3}
& @1 & @all & & \\
\midrule
I2Q              & 71.39 & 90.85 & 65.99 & 79.23\\
$+$ Enh. I2Q     & 73.03 & 92.84 & 69.16 & 80.50\\
$+$ UI2Q         & 77.09 & 93.42 & 71.72 & 82.57\\
$+$ Enh. UI2Q    & 77.96 & 93.82 & 72.96 & 83.04\\
\bottomrule
\end{tabular*}
\caption{Incremental ablation of the four curriculum subtasks: I2Q (item only),
Enhanced I2Q ($+$rationale), UI2Q ($+$user context), and Enhanced UI2Q (both).
All other components---diversity regularization, self-distillation, GRPO, the
backbone, and optimization settings---are held fixed. Soft HR is in percent.}
\label{tab:curriculum}
\end{table}

Table~\ref{tab:curriculum} isolates the curriculum by adding its four subtasks
in order while holding every other component fixed. Rationale supervision
(Enhanced I2Q) gives the largest single-step APD gain (+3.17), indicating that
reasoning helps the model explore multiple plausible search directions rather
than imitate one query; user context (UI2Q) gives the largest Soft HR@1 gain
(+4.06), confirming personalization as the dominant driver of second-hop
relevance. In total the curriculum contributes +6.57 Soft HR@1 and +6.97 APD.
This isolates the marginal contribution of each subtask, not the effect of
sequential ordering versus randomly mixed multi-task training.

\subsection{Hyper-parameter Analysis}
\begin{figure}[t]
  \centering
  \includegraphics[width=\linewidth]{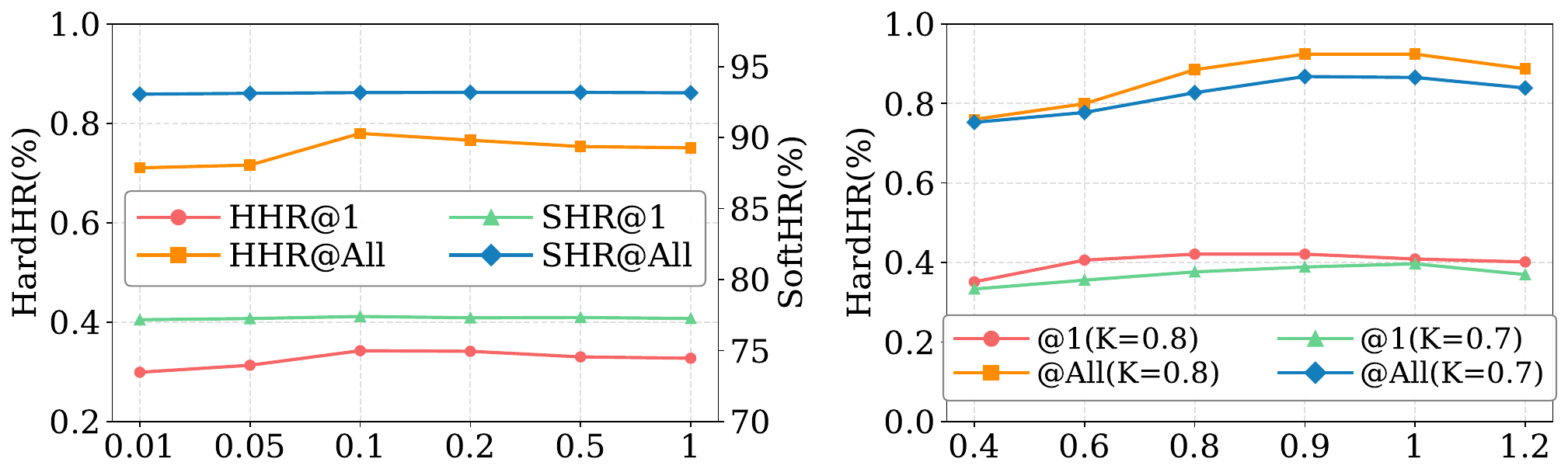}
  \vspace{-0.2em} 
  \makebox[0.5\linewidth]{\small (a)  Diversity Weight}%
  \makebox[0.5\linewidth]{\small (b) Temperature \& Top-P
}%
 \caption{Impact of hyperparameters on performance. (a) Effect of the diversity loss weight $\lambda$. (b) Effect of temperature and top-p in self-distillation.}
  \label{fig:div}
\end{figure}

Figure~\ref{fig:div} presents the sensitivity of key hyperparameters. For diversity loss weight (Figure~\ref{fig:div}(a)), performance peaks around $\lambda = 0.1$; further increasing $\lambda$ degrades relevance, suggesting overly strong regularization hurts the balance between diversity and accuracy. For self-distillation (Figure~\ref{fig:div}(b)), performance improves as temperature increases and peaks around $0.9$--$1.0$, with a larger top-$k$ consistently yielding better results.

\subsection{Online Experiments}
\label{sec:online}
We deployed EAGER as an additional query recommendation source in our production system and conducted a 9-day online A/B test with 5\% of traffic. The control uses the existing production system; the treatment augments it with EAGER-generated query suggestions while all downstream components remain identical.

As shown in Table~\ref{tab:on}, UCTR and PCTR improve by 0.83\% and 1.28\%, while DAC, Pay Count, and L2P increase by 2.98\%, 3.49\%, and 2.38\%, confirming that engagement gains translate to downstream conversion (all p < 0.05, stable throughout the test period).

\section{Conclusion}
\label{sec:conclusion}
We present EAGER, a two-stage framework that generates personalized query recommendations from clicked items, bridging item recommendation and search in e-commerce. The key insight is that intent coverage and deployment alignment are complementary objectives best addressed in sequence: the first stage expands the generation space to capture diverse user intents, while the second constrains it to meet online business requirements without sacrificing diversity. Production deployment and A/B tests confirm consistent gains in both user engagement and downstream conversion.

\section*{Limitations}
This work has several limitations. First, our method is developed and evaluated in a real-world e-commerce search scenario, where user behavior patterns, item distributions, and business rules may differ across platforms. Therefore, its generalization to other search or recommendation scenarios requires further validation. Second, PARM relies on historical exposure and click logs, which may inherit existing position bias and exposure bias from the online system. Although rule-based rewards help improve compliance, they are manually designed and may need adaptation when business requirements change. Finally, our current framework mainly focuses on query-level generation and does not jointly optimize downstream ranking or conversion objectives. Future work will explore more general reward modeling, debiased user feedback, and end-to-end optimization with downstream search modules.

\bibliography{custom}

\appendix

\label{sec:appendix}

\section{Related Work}
\paragraph{Query Recommendation}
Traditional query recommendation methods rely on heuristic rules and statistical models derived from query logs, term co-occurrence, and user behavior. They work well in narrow settings but often suffer from sparse data and struggle to adapt to changing user intents~\cite{baeza2004query, sordoni2015hierarchical}. Neural approaches improve on this by modeling sequential patterns to provide better contextual recommendations~\cite{lai2023workload}, yet they still struggle with long-tail queries. Recently, the field has moved toward generative methods using LLMs. For example, Cold-EQS~\cite{sun2026quality}  proposes an iterative reinforcement learning framework for ColdStart E-commerce Query Suggestion. PushGen~\cite{bie2026pushgen} introduces a controllable category-prompting technique to guide LLM outputs toward desired styles. GQR~\cite{min2025prompting} unifies diverse query recommendation tasks by a universal prompt framework. ReList~\citep{bi-etal-2026-relist} further recasts related-search
recommendation as a reasoning-enhanced listwise generation task, combining
chain-of-thought supervision with multi-objective reinforcement learning. All of
these methods condition on a user-typed query, whereas I2Q must generate a
personalized query set from a clicked item alone.

\graphicspath{{./image/}}
\begin{figure*}[t]
  \centering
  \includegraphics[width=\textwidth]{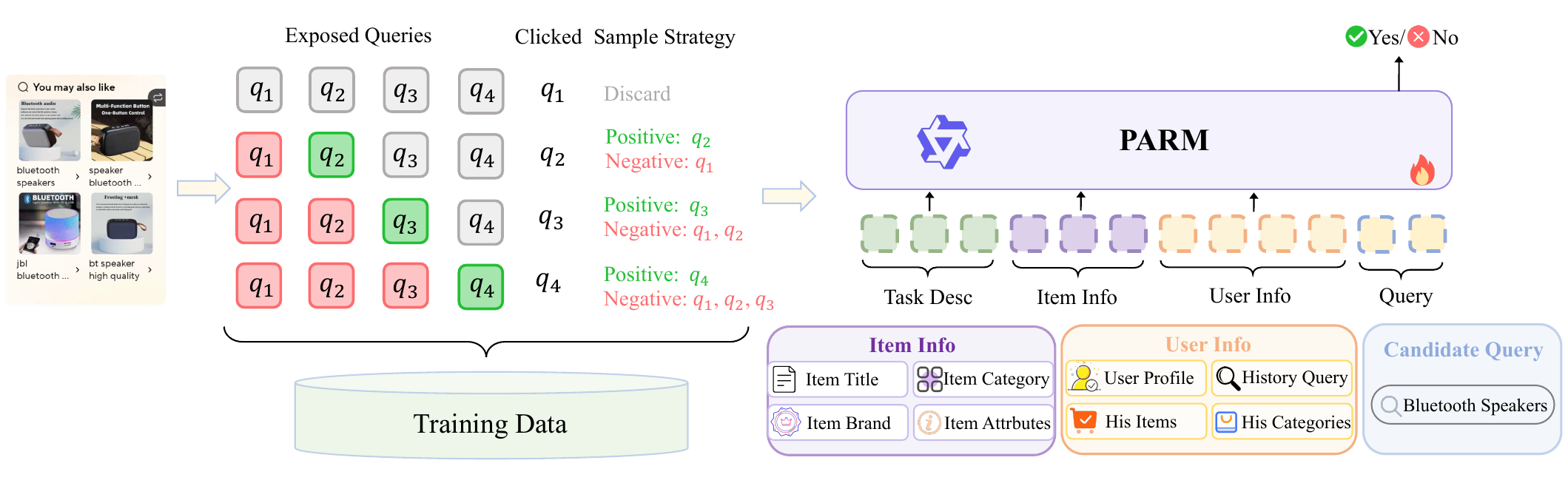}
 \caption{Training pipeline of PARM.}
  \label{fig:click_model}
\end{figure*}

\paragraph{Generative Recommendation}
Generative recommendation leverages LLMs to produce recommendations as sequences, either as semantic IDs (SIDs) or in natural language. For instance, HLLM~\cite{chen2024hllm} employs a hierarchical structure to model user interests and item representations separately. In industrial practice, the OneRec family~\cite{zhou2025onerectechnicalreport, zhou2025onerecv2technicalreport} establishes end-to-end generative frameworks that integrate reasoning and reinforcement learning for video and e-commerce recommendations. Concurrently, SynerGen~\cite{gao2025synergen} and OneSearch~\cite{chen2025onesearch} unify search and recommendation tasks under generative paradigms. However, recent work suggests natural language itself can serve as a more expressive alternative to SIDs~\cite{liu2025understanding}. Unlike SID-based approaches, EAGER operates entirely in the natural language space, directly generating contextually persuasive queries without discrete indexing, thereby fully harnessing the generative capacity of LLMs for pre-search query recommendation.

\section{PARM Training Details}
\label{sec:appendix_clk}
PARM is trained from user behavior logs in the I2Q scenario. Its goal is to estimate whether a candidate query is likely to be clicked under a given item-user context. PARM is used both to rank offline LLM-generated queries during SFT data construction and to provide the model-based reward in GRPO optimization.

\paragraph{Data Construction.}
For the I2Q task in e-commerce search, the system displays an ordered list of candidate queries. To avoid a severe imbalance between positive and negative samples, we adopt a position-aware sampling strategy based on user clicks. If the user clicks the first exposed query, we discard this event because users tend to click the first suggestion more often due to positional bias, which may distort the modeling of true user preferences. If the user clicks the second query, the clicked query is treated as a positive sample, while the first query is treated as a negative sample. Similarly, if the user clicks the third query, that query is used as the positive sample, and the first two queries are used as negative samples; if the user clicks the fourth query, the preceding three queries are used as negative samples. Queries after the clicked position are ignored, since they may not have been sufficiently examined by user.

Each training instance consists of the clicked item information, user personalization features, and a candidate query. Positive samples are assigned the output label ``yes'', while negative samples are assigned the output label ``no''.

\paragraph{Training Strategy.}
Given item information $i$, user context $u$, and a candidate query $q$, the model estimates whether the user would click the query in the current exposure context. Instead of using a conventional binary classifier, we formulate click prediction as a generative discrimination task and train the model with SFT. The input contains $(i,u,q)$, and the target output is a natural language label $a \in \{\text{yes}, \text{no}\}$.

The training objective follows the standard autoregressive negative log-likelihood:
\begin{equation}
\mathcal{L}_{\text{click}} = - \log p_\phi(a \mid i, u, q),
\end{equation}
where $\phi$ denotes the parameters of PARM. This formulation allows the model to reuse LLMs' semantic understanding and learn the matching relationships among item text, user context, and candidate queries.

\graphicspath{{./image/}}
\begin{figure*}[t]
  \centering
  \includegraphics[width=\textwidth]{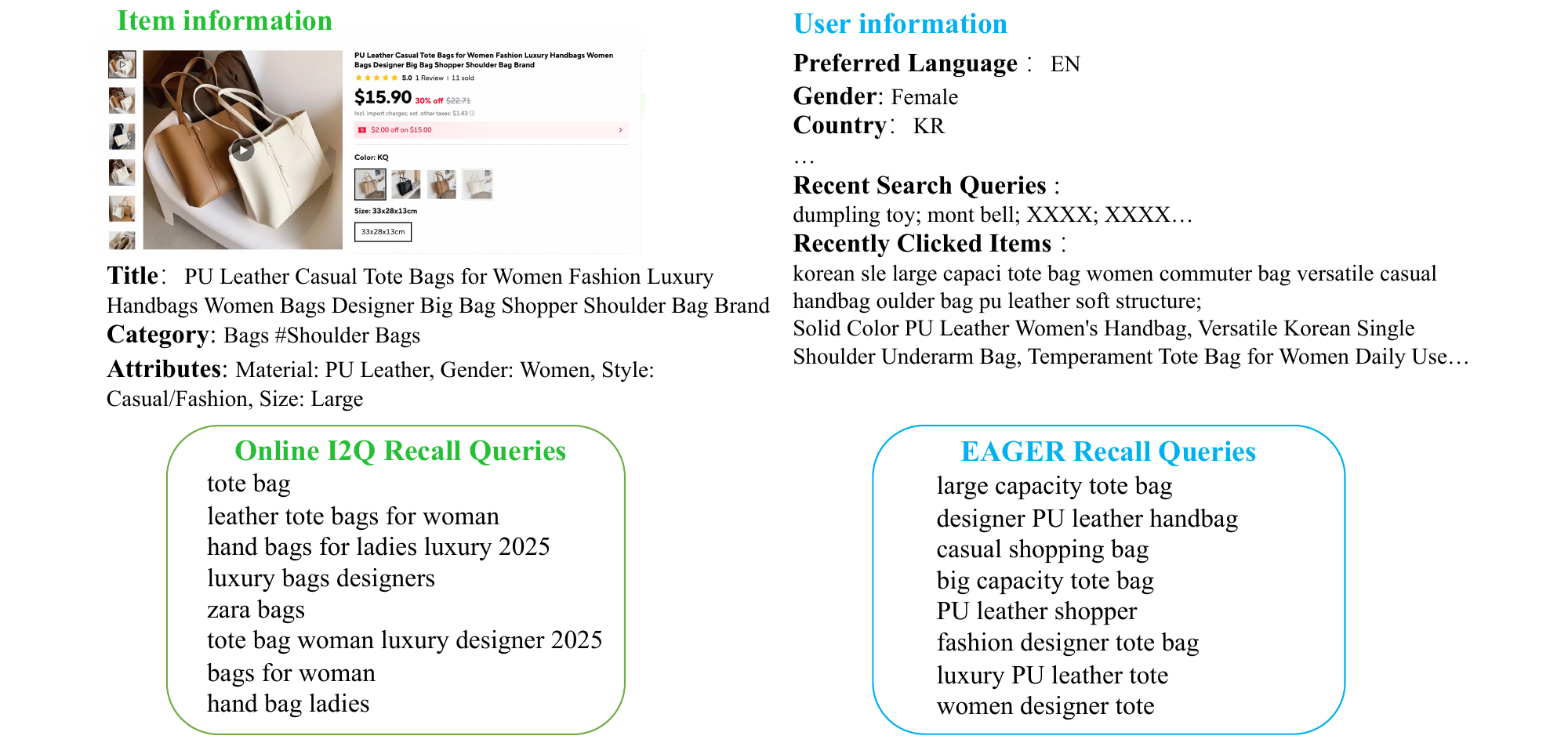}
  \caption{Qualitative comparison between the production I2Q baseline and EAGER, showing that EAGER captures fine-grained user preferences while the baseline produces generic queries}
  \label{fig:case}
\end{figure*}

\paragraph{Offline Ranking Agreement.}
To evaluate the quality of PARM as a reward signal, we measure its ranking agreement with the production ranker on a held-out sample of 210K click PVs. Specifically, for each exposure event, we compare the position rankings produced by PARM against those of the online production ranker. We report the top-$k$ match rate, defined as the proportion of events where the top-$k$ queries selected by PARM exactly match the top-$k$ set selected by the production ranker.

\begin{table}[h]
\centering
\begin{tabular}{lcccc}
\toprule
\textbf{Sample Size} & \textbf{Top-1} & \textbf{Top-2} & \textbf{Top-3} & \textbf{Top-4} \\
\midrule
210K PVs & 79.8\% & 74.1\% & 72.7\% & 72.7\% \\
\bottomrule
\end{tabular}
\caption{Ranking agreement between PARM and the production ranker.}
\label{tab:click_model}
\end{table}

PARM achieves a 79.8\% top-1 match rate with the production ranker, indicating that it can reliably identify the most attractive query for a given user-item context. The top-2 to top-4 match rates remain above 72\%, suggesting consistent alignment beyond the first position. These results confirm that PARM provides a sufficiently reliable reward signal for GRPO optimization.

\section{Rule Reward}
We design verifiable rule-based rewards to ensure that generated queries satisfy the basic requirements of the I2Q scenario. These rewards are computed offline using deterministic rules and linguistic features, without relying on online services or additional model-based evaluation.

\paragraph{Experience Compliance.}
This reward checks whether a generated query contains low-quality or experience-harmful expressions, such as ``free item'' or invalid prepositions. Queries containing such expressions are penalized to avoid unnatural or misleading search suggestions.

\paragraph{Quantity Compliance.}
This reward verifies whether the number of generated queries matches the required target number. It encourages the model to produce a stable and controllable number of candidates.

\paragraph{Lexical Overlap.}
To ensure item relevance, we compute the lexical overlap between each generated query and the item title. For queries shorter than three words, at least two words are required to match the item title; otherwise, at least 75\% of the query terms should be covered.

\paragraph{User Intent Alignment.}
This reward checks whether the generated query contains keywords from the user's historical behaviors. It encourages the model to generate personalized queries that better reflect user interests.

\paragraph{Fluency.}
We evaluate query fluency using simple linguistic features, including length constraints, token-type ratio, character validity, and bigram-level fluency. This reward encourages queries that are natural, readable, and consistent with search habits.

\paragraph{Diversity.}
To reduce redundant generation, we compute the Jaccard similarity among generated queries and reward lower similarity. This encourages the model to cover diverse search intents within the same output set.

\paragraph{Penalty Terms.}
We apply additional penalties for blacklisted or sensitive terms, including prohibited words, brand-infringing expressions, and other unsafe or low-quality content. These penalties help improve the safety and business compliance of generated queries.

\section{Evaluation Metrics}
We evaluate generated queries offline from four aspects: accuracy, relevance, diversity, and rule compliance. 

\paragraph{Hard HR@K.}
Hard HR measures the exact matching accuracy between generated queries and the ground-truth clicked query. A hit is counted if any generated query exactly matches at the string level. HardHR@1 evaluates only the first generated query, while HardHR@All evaluates the entire generated set.

\paragraph{Soft HR@K.}
Soft HR measures behavior-related relevance. A generated query is considered a soft hit if it exactly matches the ground-truth clicked query, or if at least 60\% of its terms appear in the title of the target item clicked by the user on the search result page. 
Compared with Hard HR, Soft HR better captures whether the generated query can cover the semantic intent behind the user's final clicked item.

\paragraph{Distinct-2.}
Distinct-2 evaluates lexical diversity by computing the ratio of unique 2-grams to all 2-grams in the generated query set:
\begin{equation}
\text{Distinct-2} =
\frac{|\text{Unique}(\text{2-grams}(Y))|}
{|\text{2-grams}(Y)|}.
\end{equation}
A higher Distinct-2 indicates fewer repeated expressions in the generated results.

\paragraph{APD.}
Average Pairwise Distance measures the average dissimilarity among generated queries. For each pair of generated queries, we compute their similarity $\text{sim}(q_i,q_j)$ and define the distance as $1-\text{sim}(q_i,q_j)$. APD is computed as:
\begin{equation}
\text{APD} =
\frac{1}{K(K-1)}
\sum_{i\neq j}
\left(1-\text{sim}(q_i,q_j)\right).
\end{equation}
A higher APD indicates that the generated query set covers a more diverse set of potential search intents.

\paragraph{Rule.}
The Rule metric evaluates whether the generated queries satisfy the basic display requirements and business constraints of the I2Q scenario. We reuse the verifiable rules defined in the rule-based reward, including experience compliance, quantity compliance, lexical overlap, user intent alignment, and penalty terms. The final Rule score is computed as the weighted sum of all rule scores. A higher Rule score indicates better compliance with business requirements.

\paragraph{Online Metrics.}
We evaluate online deployment from two complementary stages of the user funnel. The first stage measures user engagement, i.e., whether the generated query suggestions attract user interaction. \textbf{UCTR} (User Click-Through Rate) is the proportion of users who click at least one query suggestion among those exposed in this scenario, while \textbf{PCTR} (PV-level Click-Through Rate) is the ratio of clicked query suggestions to total query exposures, measuring per-impression attractiveness. The second stage measures downstream conversion, i.e., whether engagement translates into commercial value after the user enters search via a clicked query. \textbf{DAC} (Daily Active Customers) is the number of unique users per day who complete at least one purchase through search sessions triggered in this scenario. \textbf{Pay Count} is the total number of paid orders attributable to this scenario \textbf{L2P}(Landing-to-Payment rate) is the conversion rate from clicking a query recommendation to completing payment, capturing the end-to-end commercial efficiency of the recommended queries. Together, these metrics cover the full funnel from query exposure to final conversion.

\section{Implementation Details}
All experiments were conducted on 24 NVIDIA A100 80GB GPUs. We first performed SFT on Qwen3-1.7B~\cite{yang2025qwen3technicalreport} using the LLaMAFactory framework with full-parameter updates, a maximum sequence length of 4096, an effective batch size of 64, a learning rate of 2e-5, bf16 precision, and 1 epoch of training.

The resulting model was then used as the initial policy for reinforcement learning via GRPO in the ROLL\cite{wang2025reinforcement} framework. In the GRPO stage, each rollout response was scored by verifiable rules and PARM to obtain a reward score.

Our online inference is powered by the vLLM framework, which features a built-in multi-sampling generation mechanism. It independently generates 10 complete results, with no sharing between the candidates. The compute and VRAM usage equals the total token volume of the N generated sequences. Since they are independent, the pressure is lower than that of Beam Search. 

\begin{figure}[t]
  \centering
 \includegraphics[width=1.0\linewidth]{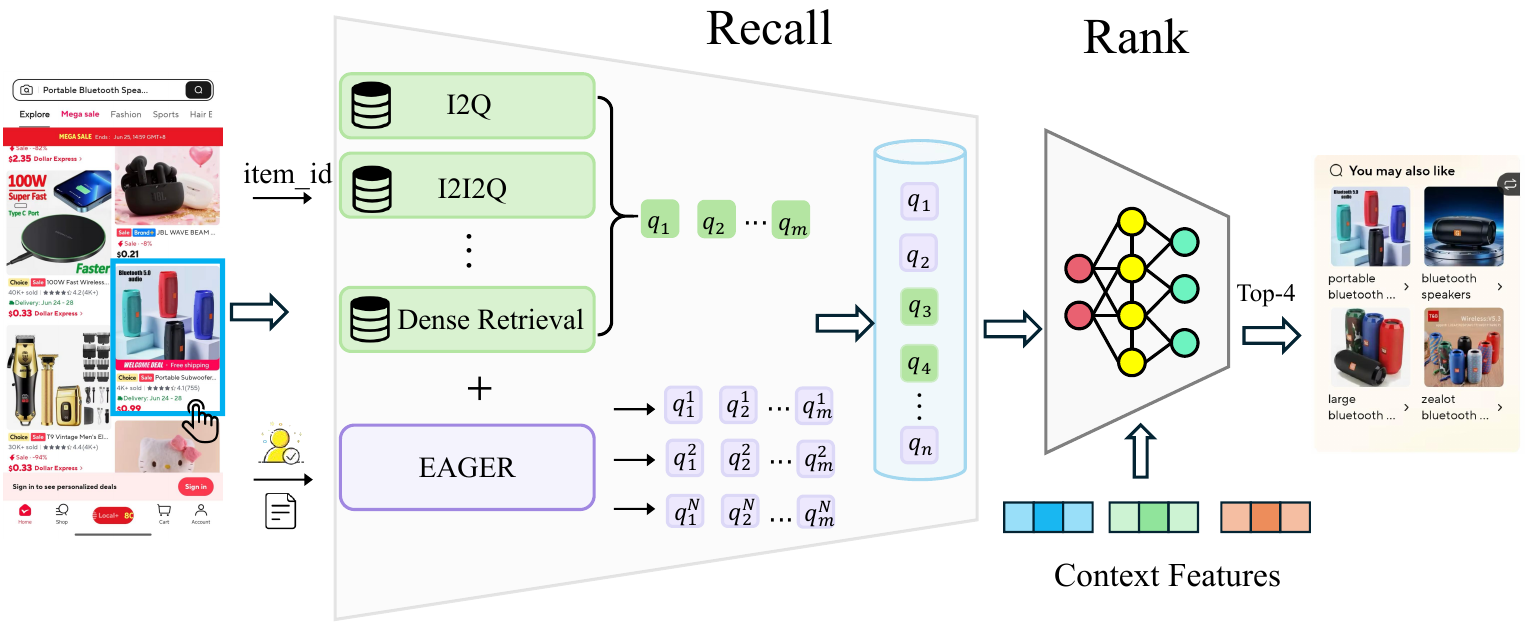}
\caption{Online serving architecture. EAGER serves as a third recall channel alongside log-based and dense retrieval; a shared ranker scores all candidates over context features and returns the top-$4$ queries.}
  \label{fig:serving}
\end{figure}

\paragraph{Online Serving Architecture.}
Figure~\ref{fig:serving} shows the deployment of EAGER in our production query recommendation system. We deploy EAGER as a third recall channel alongside two existing ones: \textbf{Log-based Retrieval}, which mines item-to-query and item-to-similar-item-to-query associations from historical click logs (I2Q and I2I2Q), and \textbf{Dense Retrieval}, which performs embedding-based approximate nearest neighbor search between item and query representations. Candidates from all channels are merged and scored by a shared ranker over context features, and the top-$K$ queries are returned to the user. EAGER complements existing channels in two ways: it expands coverage on long-tail items where co-click statistics are sparse, and it provides higher-quality candidates that compete with existing sources under the same ranker on head items.

\section{Case Study}
To validate the efficacy of EAGER in integrating user profiles with item semantics, we analyze a representative case shown in Figure~\ref{fig:case}. The interacting user is an English-speaking female from South Korea whose recent behavioral sequence indicates a converging interest in ``large capacity'' and ``designer-style'' luggage. While traditional methods yield generic queries such as ``tote bag'' or ``leather tote bags for woman,'' failing to capture fine-grained preferences for capacity and design aesthetics, EAGER leverages click history and demographic attributes to synthesize highly specific queries like ``large capacity tote bag'' and ``designer PU leather handbag.'' This demonstrates that EAGER effectively mitigates the limitations of statistical methods in personalized intent understanding, facilitating a transition from static attribute matching to dynamic interest alignment.
\section{Prompt Templates}

\begin{promptbox}{Offline Query Generation Prompt.}\small
    \textbf{Role Definition:} \\
    You are an e-commerce search query generation assistant. Your primary objective is to generate high-intent, diverse, and commercially valuable search queries to guide users' product discovery journey.

     \textbf{Task:} \\
    Generate User Search Queries for a Cross-Border E-commerce App based on the product details below.
    
    \textbf{Core Competencies:}\\
        1. Deep understanding of product info (titles, categories, attributes).\\
        2. Inferring user intent and usage scenarios.\\
        3. Generating natural, diverse queries for different shopping stages.

    \textbf{Cognitive Framework (Do not output):}\\
    \textit{Core Identity:} Direct term for the product?\\
        \textit{Function \& Scenario:} Problem solved? Usage context?\\
        \textit{Accessories \& Compatibility:} Needed add-ons? Compatible devices?\\
        \textit{Target Audience:} Typical user profile?\\
        \textit{Broad Exploration:} Search behavior if model is unknown?

    \textbf{Constraints and Rules:}\\
        \textit{Relevance:} Highly relevant to input. Do not fabricate attributes (e.g., color).\\
        \textit{Diversity:} Cover different intents (alternatives, accessories) and phrasing.\\
        \textit{Format:} Max 6 words. Use e-commerce keywords, not full sentences.\\
        \textit{Language:} Use \{\{language\}\}. Keep English brand names if applicable.\\
        \textit{Attribute Filtering:} Focus on physical traits, style, function. Ignore Origin, SKUs, specific models (e.g., M4A1), or dimensions.\\
        \textit{Avoid Over-Inference:} Be broad with ambiguous attributes (e.g., use "bed with storage" not "bed with drawers" unless specified).\\
        \textit{Safety:} No unsafe or prohibited content. 

        \textbf{Input Parameters:}\\
        Output Language: \{\{language\}\}\\
        Number of Queries: \{\{count\}\}

    \textbf{Product Information:}
    \begin{verbatim}
{
  "title": {title\},
  "category_path": {category_path},
  "attributes": {
    "features": {features}
  }
}
    \end{verbatim}
    \textbf{Output Specification:}
    You \textbf{must} return a single, valid JSON object. No text outside the JSON.
    \begin{verbatim}
{
  "language": "",
  "generated_queries": [
    { "query_text": "string" }
  ]
}
    \end{verbatim}

\end{promptbox}

\begin{promptbox}{Supervised Fine-tuning Prompt.}\small
    \textbf{Instruction:} You are a world-class cross-border e-commerce search intent expert. Your task is to generate 10 distinct, native, and high-intent search queries based on the user profile, behavior, and current product attributes.

    \medskip
    \textbf{Input:}\\
\textit{Current Product Title}:  \\
\textit{Category}: \\
\textit{Brand}: \\
\textit{User Nation/Market}:  \\
\textit{Interface Language}: \\
\textit{User Demographics}: Gender:  \\
\textit{Age}:  \\
\textit{Level}:  \\
\textit{Recent Search History}:  \\
\textit{Recent Click History}:  \\
\textit{Click Category History}: \\
\end{promptbox}


\end{document}